\documentclass[notitlepage,a4paper,aps,prd,onecolumn,superscriptaddress,nofootinbib,groupedaddress,longbibliography]{revtex4-1}
\usepackage{amsmath}
\usepackage{amsfonts}
\usepackage{amssymb}
\usepackage[utf8]{inputenc}
\usepackage[T1]{fontenc}
\usepackage[colorlinks=true]{hyperref}
\usepackage[dvipsnames]{xcolor}
\usepackage{graphicx}
\usepackage[normalem]{ulem}
\usepackage{enumerate}
\usepackage{tikz}

\usetikzlibrary{external}
\begin{document}
\title{The kinetic gas universe}

\author{Manuel Hohmann}
\email{manuel.hohmann@ut.ee}
\affiliation{Laboratory of Theoretical Physics, Institute of Physics, University of Tartu, W. Ostwaldi 1, 50411 Tartu, Estonia}

\author{Christian Pfeifer *}
\email{christian.pfeifer@ut.ee}
\affiliation{Laboratory of Theoretical Physics, Institute of Physics, University of Tartu, W. Ostwaldi 1, 50411 Tartu, Estonia}

\author{Nicoleta Voicu}
\email{nico.voicu@unitbv.ro}
\affiliation{Faculty of Mathematics and Computer Science, Transilvania University, Iuliu Maniu Str. 50, 500091 Brasov, Romania}

\begin{abstract}
A description of many-particle systems, which is more fundamental than the fluid approach, is to consider them as a kinetic gas. In this approach the dynamical variable in which the properties of the system are encoded, is the distribution of the gas particles in position and velocity space, called 1-particle distribution function (1PDF). However, when the gravitational field of a kinetic gas is derived via the Einstein-Vlasov equations, the information about the velocity distribution of the gas particles is averaged out and therefore lost. We propose to derive the gravitational field of a kinetic gas directly from its 1PDF, taking the velocity distribution fully into account. We conjecture that this refined approach could possibly account for the observed dark energy phenomenology.
\end{abstract}

\maketitle




\section{Kinetic gases instead of perfect fluids}\label{sec:1}
Numerous gravitating physical systems are described by a perfect fluid, for instance: neutron and ordinary stars, accretion discs, gas planets and last, but not least, the universe as a whole - which is our main subject of interest. Their properties are encoded in a (perfect) fluid energy momentum tensor
\begin{align}\label{eq:PF}
	T^{ab} = (p + \rho)U^a U^b + g^{ab} p\,,
\end{align}
composed of the density $\rho$, the pressure $p$, the average propagation direction of the fluid $U$ and the spacetime metric~$g$, supplemented by an equation of state relating energy density and pressure. The dynamics of the system are derived from the Euler equations $\nabla_a T^{ab} = 0$; the gravitational field is determined by the Einstein equations:
\begin{align}\label{eq:EEQ}
	G^{ab} = \frac{8\pi G}{c^4}T^{ab}\,.
\end{align}
Solving the Einstein equations in homogeneous and isotropic symmetry, and interpreting the perfect fluid energy-momentum tensor as the one describing the matter content of the whole universe, yields the standard model of cosmology and predicts the existence of $\sim 72\% $ dark energy, $\sim 23\%$ dark matter and only $\sim 5\%$ visible matter as constituents of the universe.

A more fundamental way to describe the dynamics of a fluid, is to understand the multiple particle system which constitutes the fluid as a kinetic gas \cite{Ehlers2011}. In this approach, all the information about the system is stored in a \emph{1-particle distribution function (1PDF)}, $\phi(x,\dot x)$. It encodes how many gas particles at a given spacetime point $x$ propagate on worldlines with normalized 4-velocity $\dot x = \frac{dx}{d\tau}$. The dynamics of the kinetic gas are derived from a time evolution equation along all possible particle trajectories,
\begin{align}\label{eq:1pdfev}
	\frac{d \phi}{d\tau} = \frac{\partial \phi}{\partial x^a}  \dot x^a + \frac{\partial \phi}{\partial \dot x^a}  \ddot x^a = C\,,
\end{align}
where $C$ is called the collision current of the gas. If $C=0$, the gas particles don't collide or scatter; such gases are named collisionless and \eqref{eq:1pdfev} is then called the \emph{Liouville equation}.

Usually, the gravitational field sourced by a kinetic gas is derived from the Einstein-Vlasov equations,~\cite{Andreasson:2011ng, Sarbach:2013fya, Sarbach:2013uba},
\begin{align}\label{eq:EV}
G^{ab} = \frac{8\pi G}{c^4}T^{ab} = \frac{8\pi G}{c^4} \int_{\mathcal{O}_x} \dot x^a \dot x^b \phi(x,\dot x) \Sigma_x\,.
\end{align}
Here, instead of using the velocity distribution, one obtains an energy-momentum tensor on spacetime from averaging over all normalized particle 4-velocities $\mathcal{O}_x$ at a spacetime point $x$. Thus, in the determination of the gravitational field of the kinetic gas, its velocity distribution is only taken into account on average.\\

We observe that the velocity distribution of the gas is information about gravitating matter which is not properly taken into account when deriving the gravitational field from \eqref{eq:PF} and \eqref{eq:EEQ} or from \eqref{eq:EV}. We propose to derive the gravitational field of a kinetic gas directly from the 1PDF without averaging, including the full information on the velocity distribution.\\

For the universe as a whole, our proposal means that averaging over the velocities of the cosmological kinetic gas in the description of the evolution of the universe destroys information. We conjecture that taking into account this information - the precise velocity distribution of the cosmological gas - might explain the dark energy phenomenology. Even if the amount of neglected information through averaging can locally be small, it accumulates over large, i.e.\ cosmological, distances.

To visualize the argument, we employ the following illustrative example: Consider a cloud of non-interacting dust particles, with a certain velocity distribution. On average, it propagates in the direction given by the average 4-velocity of its constituents. Over time, the cloud will evolve according to the precise distribution of velocities of the dust particles; on the fluid level, one would describe this evolution in terms of density and pressure, related by an equation of state, while, in the kinetic gas picture, all dynamical properties of the cloud are encoded in the 1PDF subject to the evolution equation.

However, in classical general relativity, when determining the gravitational field of the cloud, the velocity distribution of the particles is only effectively taken into account - in the fluid viewpoint, the velocity distribution is absorbed into the density and pressure and, in the kinetic gas case, through the energy-momentum tensor obtained via averaging. In contrast, we propose to derive the gravitational field of a kinetic gas \textit{directly} from its 1PDF, i.e.\ on a more fundamental level, without averaging and without macroscopic quantities like pressure and density. Replacing the gas cloud by the universe, i.e.\ the dust particles by galaxies, this leads to a more fundamental determination of the evolution of the universe compared to its commonly used perfect fluid description.

The question is how can we take the velocity distribution of the gas into account more accurately when determining its gravitational field: which are the appropriate equations and how are they connected to the Einstein-Vlasov or perfect fluid Einstein equations?

\section{Finsler geometry and the gravitating kinetic gas}\label{sec:2}
The 1PDF of a kinetic gas depends on the spacetime position and the possible 4-velocities of the gas particles. Technically, this makes it a function on observer space, the space of massive observer positions and velocities\footnote{We do not consider photon gases in this essay.}. In order to couple gravity directly to the 1PDF, we need a formulation of spacetime geometry and its dynamics on the same mathematical footing.

The framework which naturally does this is \emph{Finsler spacetime geometry}. It is based on the most general geometric length measure for curves $\gamma$ or, in physics language, on the most general geometric clock for observers, respectively the free massive point particle action
\begin{align}\label{eq:flnegth}
	S[\gamma] = \int d\tau\ \sqrt{|L(\gamma(\tau),\dot \gamma(\tau))|}.
\end{align}
The fundamental ingredient is a Finsler Lagrangian $L(x,\dot x)$, which is a function of positions and velocities (technically, the tangent bundle of spacetime). It is $2$-homogeneous with respect to its velocity dependence, $L(x,\lambda \dot x) = \lambda^2 L(x,\dot x),  \forall \lambda >0$, and satisfies certain smoothness properties. A Finsler spacetime $(M,L)$ is a manifold $M$ with a Finsler Lagrangian $L$, which ensures the existence of:
\begin{itemize}
	\item a well defined causal structure: a precise notion of causal (timelike and null) directions;
	\item an observer space $\mathcal{O}$: the set of normalized future pointing timelike vectors;
	\item a Finsler metric: $g^L_{ab}(x,\dot x) = \frac{1}{2}\frac{\partial}{\partial \dot x^a}\frac{\partial}{\partial \dot x^b }L$ of Lorentzian signature $(+,-,-,-)$, at least on all timelike directions;
	\item the Finsler length measure \eqref{eq:flnegth}: a geometric clock/massive point particle action.
\end{itemize}
The precise technical definition of Finsler spacetimes is given in the literature \cite{Hohmann:2018rpp,Javaloyes:2018lex}.

The geometry of a Finsler spacetime is derived from the Finsler Lagrangian $L$, similarly as spacetime geometry is derived from a pseudo-Riemannian metric in general relativity. The main difference is that geometric objects like connection coefficients and the curvature scalar depend on positions and velocities. This makes Finsler geometry the natural candidate to describe the geometry of spacetime sourced by a 1PDF.

The Finsler extension of the Einstein equations coupled to the kinetic gas can be derived from a generally covariant action principle on the observer space, see \cite{Hohmann:2019sni},
\begin{align}
	S = \int_{\mathcal{O}} \left( \frac{1}{2\kappa^2} R_0 + \phi \right)\Sigma\,,
\end{align}
where $\kappa$ is the (yet unknown) Finsler gravity coupling constant, \(R_0\) is the Finsler Ricci scalar and $\Sigma$ is the canonical volume form on $\mathcal{O}$ induced by $L$. Variation with respect to $L$ yields the field equation
\begin{align}\label{eq:fgravgas}
\mathfrak{G} := \frac{1}{2}g^{Lab}\frac{\partial}{\partial \dot x^a}\frac{\partial}{\partial \dot x^b } (R_0 L) - 3 R_0 - g^{Lab}\left(\nabla_{\delta_a}P_{b} - P_aP_b + \frac{\partial}{\partial \dot x^a}(\nabla P_b)\right) = -\kappa^2\phi\,.
\end{align}
This is a scalar equation on the observer space whose left hand side \(\mathfrak{G}\) describes the Finsler geometry of spacetime sourced by the 1PDF of the gas on its right hand side. The traced Landsberg tensors $P_a=P^d{}_{da}$ measure how the departure of the spacetime geometry from the usual general relativistic setup varies in position and direction. More technically, the $\dot x$ derivative of the Finsler metric, usually called the Cartan tensor, measures the departure of a given Finsler geometry from (pseudo)-Riemannian geometry; the Landsberg tensor then is given by the so-called dynamical covariant derivative of this quantity  $P^a{}_{bc}=\frac{1}{2}g^{Lad}\nabla \left(\frac{\partial}{\partial \dot x^b}g^L_{cd}\right)$. For further mathematical details about the Finsler geometric objects appearing we refer to the articles \cite{Hohmann:2019sni,Javaloyes:2018lex} and the books~\cite{Shen,Bucataru} as well as references therein.

We can relate \eqref{eq:fgravgas} to an \emph{effective} gravitational field density equation on spacetime via averaging
\begin{align}
	\underbrace{\int_{\mathcal{O}_x} \dot x^c \dot x^d \left(\frac{1}{2}g^{Lab}\frac{\partial}{\partial \dot x^a}\frac{\partial}{\partial \dot x^b } (R_0 L) - 3 R_0 - g^{Lab}\left(\nabla_{\delta_a}P_{b} - P_aP_b + \frac{\partial}{\partial \dot x^a}(\nabla P_b)\right)\right)  \Sigma_x}_{\mathfrak{G}^{cd}} =  -\kappa^2 \underbrace{\int_{\mathcal{O}_x} \dot x^c \dot x^d \phi \Sigma_x}_{\mathfrak{T}^{cd}}\,,
\end{align}
which is reminiscent of the Einstein-Vlasov equations. It is under investigation under which conditions the tensor density \(\mathfrak{G}^{cd}\) can be interpreted as an improved Einstein tensor, containing also dark energy. In these cases we obtain a correction to the Einstein-Vlasov equations in the effective description of a gravitating kinetic gas.

\section{Kinetic gas cosmology}\label{sec:3}
As a simple example, and a main motivation for our approach, we apply the kinetic gas model to cosmology. The cosmological principle, i.e.\ spatial homogeneity and isotropy, implies that the Finsler Lagrangian must be of the form~\cite{Pfeifer:2011xi}
\begin{equation}
	L = L(t, \dot{t}, w)\,, \quad w^2 = \frac{\dot r^2}{1- kr^2} + r^2 (\dot \theta^2 + \sin^2\theta \dot\phi^2)\,.
\end{equation}
For the gravity field equation~\eqref{eq:fgravgas} on the observer space, this implies \(\mathfrak{G} = \mathfrak{G}(t, s)\) and \(\phi = \phi(t, s)\), where \(s = w / \dot{t}\) measures the speed of gas particles relative to their averaged background motion. Comparing this to homogeneous and isotropic cosmology in general relativity (which inevitably leads to a Friedmann-Lemaitre-Robertson-Walker metric described by one function of \(t\) only), we see that the kinetic gas approach allows for a much larger class of geometries, for example containing two functions of $t$
\begin{align}
	L = (\dot t^2 - a(t)^2 w^2)e^{ \frac{ b(t)^2 w^2}{\dot t^2-a(t)^2 w^2}}\,,
\end{align}
and may thus potentially accommodate for a richer phenomenology. Finding solutions to the cosmological field equation sourced by the 1PDF of the matter content of the universe, and studying the resulting dynamics - in particular the possibility of modelling the accelerating expansion of the universe - is thus a primary goal. As a first step, one may search for solutions among the recently found class of cosmological Finsler Berwald spacetimes~\cite{Hohmann:2020mgs}, for which the field equations simplify significantly \cite{Fuster:2018djw}.

\begin{acknowledgments}
C.P. and M.H. were supported by the Estonian Ministry for Education and Science through the Personal Research Funding Grants PSG489 (C.P.) and PRG356 (M.H.), as well as the European Regional Development Fund through the Center of Excellence TK133 ``The Dark Side of the Universe''. The authors would like to acknowledge networking support by the COST Actions CANTATA (CA15117) and QGMM (CA18108), supported by COST (European Cooperation in Science and Technology).
\end{acknowledgments}

\bibliographystyle{utphys}
\bibliography{KGU}

\end{document}